\documentclass[preprintnumbers,article,amsmath,amssymb,floatfix,10pt,prd,twocolumn,
superscriptaddress,nofootinbib]{revtex4-2}
\usepackage{bm}
\usepackage{amsfonts}
\usepackage{latexsym}
\usepackage[latin1]{inputenc}
\usepackage{graphicx}
\usepackage{amsmath}
\usepackage{palatino}
\usepackage{mathpazo}
\usepackage{textcomp}
\usepackage{float}
\usepackage{booktabs}
\usepackage{dcolumn}
\usepackage{ragged2e}
\usepackage{hyperref}
\hypersetup{colorlinks,citecolor=blue}
\hypersetup{colorlinks=true,linkcolor=red,filecolor=magenta,    urlcolor=blue}
\usepackage{amsmath}
\usepackage{xcolor}
\usepackage{orcidlink}
\usepackage{epsfig}
\usepackage[caption=false]{subfig}
\usepackage{commath}
\usepackage{tabularx}
\usepackage{multirow}
\usepackage[mathscr]{euscript}

\newcommand{\be}{\begin{equation}}
\newcommand{\ee}{\end{equation}}
\newcommand{\bea}{\begin{eqnarray}}
\newcommand{\eea}{\end{eqnarray}}
\newcommand{\eeas}{\end{eqnarray*}}
\newcommand{\beas}{\begin{eqnarray*}}

\def\jnl@style{\it}
\def\aaref@jnl#1{{\jnl@style#1}}

\def\aaref@jnl#1{{\jnl@style#1}}

\def\aj{\aaref@jnl{AJ}}                   
\def\apj{\aaref@jnl{ApJ}}                 
\def\apjl{\aaref@jnl{ApJ}}                
\def\apjs{\aaref@jnl{ApJS}}               
\def\apss{\aaref@jnl{Ap\&SS}}             
\def\aap{\aaref@jnl{A\&A}}                
\def\aapr{\aaref@jnl{A\&A~Rev.}}          
\def\aaps{\aaref@jnl{A\&AS}}              
\def\mnras{\aaref@jnl{Mon.~Not.~Roy.~Astron.~Soc.}}             
\def\prd{\aaref@jnl{Phys.~Rev.~D}}        
\def\prc{\aaref@jnl{Phys.~Rev.~C}}  
\def\prl{\aaref@jnl{Phys.~Rev.~Lett.}}    
\def\qjras{\aaref@jnl{QJRAS}}             
\def\skytel{\aaref@jnl{S\&T}}             
\def\ssr{\aaref@jnl{Space~Sci.~Rev.}}     
\def\zap{\aaref@jnl{ZAp}}                 
\def\nat{\aaref@jnl{Nature}}              
\def\aplett{\aaref@jnl{Astrophys.~Lett.}} 
\def\apspr{\aaref@jnl{Astrophys.~Space~Phys.~Res.}} 
\def\physrep{\aaref@jnl{Phys.~Rep.}}      
\def\physscr{\aaref@jnl{Phys.~Scr}}       
\def\commat{\aaref@jnl{Comm.~Math.~Phys.}}              
\def\science{\aaref@jnl{Science}}               
\def\cqg{\aaref@jnl{Classical Quant.~Grav.}}            
\def\jpcs{\aaref@jnl{JPCS}}                                     
\def\ijmpd{\aaref@jnl{Int.~J.~Mod.~Phys.~D}}                    
\def\grg{\aaref@jnl{Gen.~Relat.~Gravit.}}               
\def\rpp{\aaref@jnl{Rep.~Prog.~Phys.}}          
\def\npa{\aaref@jnl{Nucl.~Phys.~A}}        
\def\lrr{\aaref@jnl{Living Rev.~Rel.}}                   
\def\jcap{\aaref@jnl{J.~Cosmology Astropart.~Phys.}}    
\def\rmp{\aaref@jnl{Rev.~Mod.~Phys.}}   
\def\epjc{\aaref@jnl{Eur.~Phys.~J.~C}} 
\def\plb{\aaref@jnl{~Phy.~Lett.~B}} 
\def\mpla{\aaref@jnl{Mod.~Phy.~Lett.~A}} 
\def\arxiv{\aaref@jnl{arxiv.org}}

\allowdisplaybreaks[1]

\begin{document}
\color{black}       
\title{Exploring Phase Space Trajectories in $\Lambda$CDM Cosmology with $f(G)$ Gravity Modifications}
\author{N. Myrzakulov\orcidlink{0000-0001-8691-9939}}\email{nmyrzakulov@gmail.com}
\affiliation{L. N. Gumilyov Eurasian National University, Astana 010008, Kazakhstan}
\author{Anirudh Pradhan\orcidlink{0000-0002-1932-8431}}
\email{pradhan.anirudh@gmail.com}
\affiliation{Centre for Cosmology, Astrophysics and Space Science (CCASS),\\
	GLA University, Mathura-281406, U.P., India.}
\author{Archana Dixit\orcidlink{0000-0003-4285-4162}}\email{archana.ibs.maths@gmail.com}
\affiliation{Department of Mathematics, Gurugram University, Gurugram, Haryana, India.}
\author{S. H. Shekh\orcidlink{0000-0003-4545-1975}}
\email{da\_salim@rediff.com}
\affiliation{Department of Mathematics, S.P.M. Science and Gilani Arts, Commerce College, Ghatanji, Yavatmal, Maharashtra 445301, India}

\begin{abstract}
In this work, the cosmic solutions, particularly the well-known $\Lambda$CDM model, are investigated in the framework of the Gauss-Bonnet gravity, where the gravitational action incorporates the Gauss-Bonnet invariant function. We utilize a specialized formulation of the deceleration parameter in terms of the Hubble parameter $H$, given by $q = -1 - \frac{\dot{H}}{H^2}$, to solve the field equations. To identify the appropriate model parameters, we align them to the most recent observational datasets, which include 31 data points from the Cosmic Chronometers, Pantheon+, and BAO datasets. The physical characteristics of the cosmographic parameters, such as pressure and energy density, that correlate to the limited values of the model parameters, are examined. The evolution of the deceleration parameter suggests a transition from a decelerated to an accelerated phase of the universe. Additionally, we examine the stability of the assumed model and provide an explanation for late-time acceleration using the energy conditions. The behavior of the equation of state parameter has been analyzed through dynamical variables by constraining various parameters in light of the recent observational data. This study has resulted in a quintessence-like evolution. 
\end{abstract}


\maketitle

\section{Introduction}\label{I}

Observations of Supernovae Type Ia (SNe Ia) have not only revolutionized the field of relativistic astrophysics and cosmology but also led to several intriguing discoveries and implications \cite{Riess/1998,Riess/2004,Perlmutter/1999}. These supernovae, known for their consistent peak luminosity, serve as crucial cosmological distance indicators. By comparing their observed brightness with their intrinsic luminosity, astronomers can determine their distance from Earth and, consequently, measure the expansion rate of the universe. One of the most profound discoveries from these observations is the accelerated expansion of the universe, indicating the presence of a mysterious entity known as dark energy (DE). DE constitutes a significant fraction of the universe's energy density and is responsible for the observed acceleration, a phenomenon that challenges our understanding of fundamental physics. Moreover, SNe Ia observations have provided useful views into the nature of dark energy and the cosmic expansion history. These observations, combined with other cosmological probes such as the cosmic microwave background radiation (CMBR) \cite{Komatsu/2011,Huang/2006} and baryon acoustic oscillations (BAO) \cite{D.J.,W.J.}, have led to the development of the standard cosmological model, known as the Lambda Cold Dark Matter ($\Lambda$CDM) model. This model successfully explains a wide range of cosmological observations and provides a framework for understanding the evolution and structure of the universe on large scales \cite{Steinhardt/2011}. Observational data from various sources, including the Planck satellite \cite{Planck/2020}, indicate that the energy fractions of baryonic matter (BM) and DM are approximately 5\% and 25\%, respectively. In contrast, DE comprises the remaining 70\%.

The $\Lambda$CDM cosmological model, while successful in explaining a wide range of observations, is not without its challenges. One of the significant issues it faces is the fine-tuning problem \cite{Weinberg/1989}, which arises from the need to precisely balance the cosmological constant $\Lambda$ with the energy density of matter in the universe. This fine-tuning is necessary to achieve the observed flatness and accelerated expansion of the universe. The extremely small but non-zero value of the cosmological constant required for this balance is considered unnatural and has led to discussions about the anthropic principle and the multiverse hypothesis. Another challenge is the coincidence problem \cite{Steinhardt/1999}, which questions why the energy densities of DM, DE, and BM are of the same order of magnitude in the present epoch, despite evolving differently over cosmic time. This apparent coincidence is puzzling, as there is no underlying theoretical reason why these densities should be comparable at the present epoch. Addressing these challenges has led to the exploration of alternative cosmological models, such as modified gravity theories (MGTs), scalar-tensor theories, and theories involving extra dimensions, among others \cite{Clifton}.

MGTs have emerged as a promising avenue for explaining the cosmic accelerated expansion, offering alternatives to the standard cosmological model. These theories involve modifying the standard Einstein-Hilbert action by replacing the Ricci scalar with generic functions, such as $f(R)$ or combinations of scalar and tensorial curvature invariants. This approach has gained traction among relativistic astrophysicists as a way to address the shortcomings of the standard $\Lambda$CDM model. The introduction of MGTs has led to a standard terminology that provides a framework for exploring the reasons behind cosmic acceleration. For more reviews on modified gravity, check out \cite{R1,R2,R3,R4,R5,R6,R7,R8,R9,R10,R11,R12,R13}. The initial evidence for an accelerating universe in $f(R)$ gravity was proposed by Nojiri and Odintsov \cite{R14}. Subsequent studies have revealed intriguing findings regarding the influence of dark source terms on the dynamic evolution of stellar systems in various gravitational theories, including $f(R)$ \cite{R15,R16,R17}, $f(R,T)$ \cite{R18} (where $T$ is the trace of the energy-momentum tensor), and $f(R,T,R_{\mu\nu}T^{\mu\nu})$ gravity \cite{R18,R19,R20,R21}.

In addition to the MGTs discussed in the literature, the Gauss-Bonnet (GB) gravity has garnered significant attention. Referred to as $f(G)$ gravity, it is defined by $G = R - 4{R_{\mu \nu }}{R^{\mu \nu }} + {R_{\mu \nu
\alpha \beta }}{R^{\mu \nu \alpha \beta }}$, a topological invariant in four-dimensional spacetime. This theory's equation of motion necessitates coupling with a scalar field or that $f(G)$ be an arbitrary function of $G$ \cite{G9,G10,G11,G12,G13,G14,G15,G16,G17,G18,G19}. This MGT offers valuable insights into several cosmological phenomena. It can contribute to the study of the inflationary era, the transition from deceleration to acceleration, and successfully passes tests based on solar system experiments. Moreover, various viable $f(G)$ models demonstrate the ability to cross the phantom divide line \cite{G11,G12}. Compared to the $f(R)$ gravity, the GB gravity appears to be less constrained \cite{G13}. The $f(G)$ gravity also serves as an effective framework for investigating cosmic issues as an alternative to the DE \cite{G14}. In addition, the $f(G)$ gravity is instrumental in studying finite-time future singularities and the late-time acceleration of the universe \cite{G15,G16}. In the same way, the transition from cosmic acceleration to the matter-dominated era can also be elucidated by various viable models within the $f(G)$ gravity \cite{G13,G14}. Several consistent $f(G)$ models have been proposed to satisfy specific constraints from the solar system \cite{G13,G14}, detailed in \cite{G17}. Furthermore, additional constraints on $f(G)$ models may emerge from the analysis of energy conditions \cite{G18,G19,G20}. Nojiri et al. \cite{bin4} have explored fundamental cosmic issues such as inflation, late-time acceleration, and bouncing cosmology, suggesting that modified gravity theories like $f(R)$, $f(G)$, and $f(\mathcal{T})$ (where $\mathcal{T}$ is the torsion scalar) could serve as effective mathematical tools for understanding the universe's evolution.

In this study, we focus on the reconstruction method applied to the Friedman-Robertson-Walker (FRW) universe within the framework of the $f(G)$ gravity theory. The approach involves using a parametric form for the scale factor to derive exact solutions to the field equations. The structure of this paper is organized as follows: Sec.II introduces the field equations of the $f(G)$ gravity. Observational constraint, various data and methology  are discussed in Sec.III. Results of Obserbational constraint and physical behaviour of the parameter are  discussed in Sec.IV. In Sec. V we have discussed the energy condition. The behaviour of deceleration parameter are anlysed in Sec.VI. The State finder analysis are discussed in Sec. VII and concluding remarks are mentioned in Sec. VIII.
\section{Field Equations in $f(G)$ gravity}\label{sec2}
 
Let us start from a brief review of the $f(G)$ gravity and formulation of its field equations. 
The modified Gauss-Bonnet gravity is described by the action
\begin{equation}\label{e1}
S=\frac{1}{2k} \int d^{4} x \sqrt{-g} \left[R+f(G)\right]+S_{m} (g^{\mu \nu } ,\psi ). 
\end{equation}
Here $k$ - the coupling constant, $g$ - the determinant of the metric tensor $g_{\mu \nu } $, and $S_{m} (g^{\mu \nu } ,\psi )$ is the matter action, where matter is minimally coupled to the metric tensor, with $\psi$ being the matter fields. This coupling of matter to the metric tensor implies the $f\mathrm{(}G\mathrm{)}$ gravity to be a purely metric theory of gravity where the Gauss--Bonnet invariant \textit{G} is defined as

\begin{equation}\label{e2}
G=R^{2} -4R_{\mu \nu } R^{\mu \nu } +R_{\mu \nu \sigma \rho } R^{\mu \nu \sigma \rho } . 
\end{equation}

Here \textit{R} is the Ricci scalar, and $R_{\mu \nu } $, $R_{\mu \nu \sigma \rho } $ are the Ricci and Riemann tensors, respectively.\\
The variation of the action (\ref{e1}) in terms of the metric tensor yields the gravitational field equations:
\begin{widetext}
\[R_{\mu \nu } -\frac{1}{2} Rg_{\mu \nu } +8\left[R_{\mu \nu \sigma \rho } +R_{\rho \nu } g_{\sigma \mu } -R_{\rho \sigma } g_{\nu \mu } -R_{\mu \nu } g_{\sigma \rho } +R_{\mu \sigma } g_{\nu \rho } +\frac{1}{2} R(R_{\mu \nu } g_{\sigma \rho } -R_{\rho \sigma } g_{\nu \mu } )\right]\times \] 
\begin{equation} \label{e3} 
\nabla ^{\rho } \nabla ^{\sigma } F+(Gf_{G} -f)g_{\mu \nu } =kT_{\mu \nu }. 
\end{equation} 
\end{widetext}
Here $\nabla ^{\rho }$ is the covariant derivative, whereas $f_{G} $ is the derivative of $f$ with respect to $G$.
The next step is to consider the spatially homogeneous and isotropic Friedman-Robertson-Walker (FRW) metric
\begin{equation} \label{e4} 
ds^{2} =dt^{2} -a^{2} (t)\left[\frac{dr^{2} }{1-kr^{2} } +r^{2} d\theta ^{2} +r^{2} \sin ^{2} \theta {\rm \; }d\phi ^{2} \right], 
\end{equation} 
where $a$ is the scale factor of the universe.

The source of the energy-momentum tensor for perfect fluid in the universe is 
\begin{equation} \label{e5} 
T_{\mu \nu } =(\rho +p)u_{\mu } u_{\nu } -pg_{\mu \nu } , 
\end{equation} 
where $u_{\mu } $ is the four-velocity, $\rho $ and $p$ are energy density and pressure of the universe respectively. Using the field equations, the metric and the source, Eqs.~(3)-(5), we obtain the following field equations 
\begin{equation} \label{e6} 
6\frac{\dot{a}^{2} }{a^{2} } +24\frac{\dot{a}^{3} }{a^{3} } \dot{f}_{G} -Gf_{G} +f=2k^{2} \rho , 
\end{equation} 
\begin{equation} \label{e7} 
-4\frac{\ddot{a}}{a} -2\frac{\dot{a}^{2} }{a^{2} } -8\frac{\dot{a}^{2} }{a^{2} } \ddot{f}_{G} -16\frac{\dot{a}\ddot{a}}{a^{2} } \dot{f}_{G} +Gf_{G} -f=2k^{2} p. 
\end{equation} 
Here dots represent differentiation with respect to cosmic time. 
The Gauss--Bonnet invariant is then found to be
\begin{equation} \label{e8} 
G=24\frac{\dot{a}^{2} \ddot{a}}{a^{3} } . 
\end{equation} 
With respect to the Hubble parameter $H=\frac{\dot{a}}{a} $, the field equations (6), (7) and the Gauss--Bonnet invariant (8) can be rewritten as
\begin{equation} \label{e9} 
2k^{2} \rho =6H^{2} +24H^{3} \dot{f}_{G} -Gf_{G} +f, 
\end{equation} 
\begin{equation} \label{e10} 
2k^{2} p=-4\dot{H}-6H^{2} -8H^{2} \ddot{f}_{G} -16H\left(\dot{H}+H^{2} \right)\dot{f}_{G} +Gf_{G} -f, 
\end{equation} 
\begin{equation} \label{e11} 
G=24H^{2} \left(\dot{H}+H^{2} \right). 
\end{equation} 

The contribution of the Gauss-Bonnet term to energy density and pressure in the Friedman's equation are obtained by replacing $\rho $ and$p$ to $\rho +\rho _{G} $ and $p+p_{G}$, respectively. Here $\rho _{G} $ and $p_{G} $ are the corresponding Gauss--Bonnet energy density and pressure. 
Here it is implied that the total energy density and pressure in the modified field equations can not be dark energy, and the fluid is not viscous.
Therefore, the respective components of these equations are
\begin{equation} \label{e13} 
\rho +\rho _{G} =\frac{3H^{2} }{k^{2} } , 
\end{equation} 
\begin{equation} \label{e14} 
p+p_{G} =-3q\frac{H^{2} }{k^{2} } , 
\end{equation} 
where $q$ is the deceleration parameter. 
Hence, to construct the Gauss-Bonnet contribution to energy density and pressure, the field equations (9) and (10) are rewritten as
\begin{equation} \label{e17} 
\rho _{G} =\frac{1}{2k^{2} } \left\{6H^{2} +24H^{3} \dot{f}_{G} -Gf_{G} +f\right\}, 
\end{equation} 
\begin{equation} \label{e18} 
p_{G} =\frac{1}{2k^{2} } \left\{-4\dot{H}-6H^{2} -8H^{2} \ddot{f}_{G} -16H\left(\dot{H}+H^{2} \right)\dot{f}_{G} +Gf_{G} -f\right\}, 
\end{equation} 
The equation of state parameter, obtained from Eqs. (14) and (15), is
\begin{equation} \label{e22} 
\omega _{G} =-1-\frac{1}{(2k^2)\rho_G}\left\{4\dot{H}+16H\dot{H}\dot{f}_G-8H^3\dot{f}_G+8H^2\ddot{f}_G \right\}. 
\end{equation} 
For the de-Sitter universe, i.e. $\dot{H} \rightarrow 0$ and $H^2\ddot{f}_G \rightarrow 0$, we obtain $\omega _{G} =-1$, implying the contribution of the Gauss-Bonnet term to the flat FRW model is just like as cosmological constant. Next we have to discuss the cosmological behavior of the model. For that we consider the $f(G)$ model of the form $f(G)=\alpha G^{m+1}$, where both $m$ and $\alpha$ are the free model parameters.  Since, $m = -1$, then $f(G) = \alpha $ which is a constant function. Hence we take $m \neq -1$ throughout the analysis. Also, to investigate how the cosmic parameters evolve, we now need one more ansatz. \\
Since it typically takes into account parametrizations of any kinematic parameters, including the Hubble, deceleration, and jerk parameters, and provides the required additional equation, the methodology is widely known as the model-independent method of studying cosmological models. Determining the nature of the expansion rate of the Universe requires a precise parameterization of the Hubble parameter. Several approaches, including the problem of all-time decelerating expansion, the initial singularity problem, the horizon problem, and the Hubble tension, have been extensively studied in the literature to define challenges with cosmological questions. In general, this study's normalized Hubble parameter is regarded as parametrized. We take the widely accepted Hubble parameter in terms of the redshift $z$:

\begin{equation}
H(z)=\frac{H_0}{\sqrt{2}}\left[1+(1+z)^{(2(1+\alpha)) }   \right]^{\frac{1}{2}},
\end{equation}
where $H_0$ represents the present value of Hubble's parameter $H(z)$ and $\alpha$ is any arbitrary free model parameter. The above equation yields the cosmic scale factor in cosmic time as
$a=\left[\text{sinh}\left((1+\alpha)\beta t\right)^{\frac{1}{2}}+c\right]^{\frac{1}{1+\alpha}}$, where $\beta$ and $c$ are the constants of integration. Next, in this research,
we have constrained both $H_0$ and $\alpha$ by fitting the experimental data and their measurements as below.

\section{Observational constraints}
This section presents the latest observational data and the techniques used for their analysis. It is well established that datasets from the Cosmic Microwave Background (CMB), Baryon Acoustic Oscillations (BAO), and Type Ia Supernovae are effective in constraining cosmological models. Observing the universe dynamics, a direct and model-independent view are studied, particularly through the $H(z)$ datasets, which offers a detailed depiction of the universe expansion over time \cite{ref5}. The ages of the oldest and most passively evolving galaxies facilitate direct measurements of $H(z)$ across different redshifts, thus creating a unique standard probe known as "standard clocks" in cosmology.

There are two main methods for determining $H(z)$: the radial BAO size method and the galaxy differential age method, also referred to as the Cosmic Chronometer (CC) method. In this study, we employ recent data from both the CC and BAO methods, along with Supernovae observations, to constrain the parameters of our cosmological model.  
Additionally, we use the Markov Chain Monte Carlo (MCMC) method, Bayesian analysis, and the likelihood function from the emcee Python library to analyze the data.

In this section, we utilize a range of datasets, such as observational data from the Hubble Space Telescope and distance modulus data from Type Ia Supernovae (SNe Ia), to ascertain the parameter values of our model that effectively characterize various cosmic epochs. For determining the values of 
$H_{0}$ and $\alpha$ and the model parameter, we incorporate 31 data points from various datasets and 1701 data points from the Pantheon+ sample.

\subsection{Methodology}

Markov Chain Monte Carlo (MCMC) is an  prominent  statistical approach in cosmology, used to investigate complex model parameter spaces and derive probability distributions for cosmological parameters. The MCMC approach is particularly helpful when there is a large parameter space and the likelihood function is not Gaussian or not linear. 
The main idea of using the MCMC method is to construct a Markov chain that tests the parameter space of a model based on a probability distribution \cite{ref1,ref2,ref3}. This chain consists of a series of parameter values, each generated from the previous one using transition rules impacted by a proposal distribution. It proposes new parameter values that may or may not be accepted based on their posterior probability, provided the data and the prior probability distribution \cite{ref4}.


\subsection{Cosmic chronometers Dataset}
Cosmic chronometers are astronomical objects that offer crucial data for
understanding the universe's expansion history. This data is derived 
from measuring the ages of galaxies at various redshifts. 
Elliptical galaxies are commonly used as cosmic chronometers
due to their relatively simple stellar populations and their early 
formation in the universe history \cite{ref6}.
However, they are more difficult to observe and analyze compared to other galaxies, making their use a subject of ongoing research \cite{ref7}. Astronomers use huge telescopes and spectrographs to measure the spectra and colors of these galaxies precisely in order to collect data for cosmic chronometry \cite{ref8}. 
The ages of the galaxies and the history of the universe expansion are then ascertained through the application of sophisticated statistical techniques. 
 
The data from these chronometers is essential for understanding the universe and the fundamental laws of physics. Using the differential age (DA) method, 31 Hubble data points within the redshift range $0.07 \leq z \leq 1.965$
have been analyzed \cite{ref9}. This method determines the universe expansion rate at a given redshift, allowing $H(z)$ to
be calculated using $H(z)=\frac{1}{1+z} \frac{dz}{dt}$. To obtain the mean values of the model parameters $H_{0}$ and $\alpha$, 
the chi-square ($\chi^{2}$) for the Hubble datasets can be expressed accordingly,
\begin{equation}\label{e34} \chi_{H(z)}^2=\sum_{n=1}^{31}\frac{\left[H_{th}\left(z_n,X_i\ \right)-H_{obs}\left(z_n\right)\right]^2}{\sigma_{H_{obs}\left(z_n\right)}^2},\ \ \ \ \
	\end{equation}
	where $H_{th}\left(z_n, X_i\ \right)$ is the theoretical value of the Hubble parameter which is obtained from Eq.~\textcolor{red}{(17)} in which $X_i=\left\{ H_0,\alpha\right\}$ is the set of model parameters with Hubble constant $H_0$, and $H_{obs}\left(z_n\right)$  is the observed value of  Hubble parameter. $\sigma_{H_{obs}\left(z_n\right)}^2$ is the standard error in the observed Hubble parameter measurements.  \\
\subsection{Pantheon$+$}
The Pantheon+ sample is an extension of the original Pantheon datasets, which is one of the largest and most comprehensive compilations of Type Ia supernovae (SNe Ia) data. Type Ia supernovae are used as "standard candles" because of their consistent intrinsic brightness, making them reliable distance indicators across cosmological scales \cite{ref10}. It includes over 1,500 SNe Ia, covering a redshift range from about 0.01 to 2.26. This broad range allows for detailed analysis of the universe expansion history. The Pantheon+ sample benefits from refined calibration techniques, reducing systematic uncertainties and increasing the precision of cosmological measurements. In the same context the SH0ES (Supernovae, $H_{0}$, for the Equation of State) project focuses on directly measuring the Hubble constant using a local 
distance ladder approach \cite{ref11}. 

The SH0ES team's work is critical in determining $H_{0}$ by comparing 
the redshifts and distances of supernovae, providing a direct
measurement of the local expansion rate of the universe. The SH0ES measurements of $H_{0}$ (around 73-74 km/s/Mpc) are notably higher than the value inferred from the Cosmic Microwave Background (CMB) data (around 67-68 km/s/Mpc) observed by the Planck satellite. This discrepancy, known as the "Hubble tension," is a significant topic in cosmological research, prompting investigations into potential new physics or systematic errors \cite{ref12}.

We utilize the Pantheon+SH0ES datasets, available from the GitHub repository. This collection includes 1701 light curve measurements from 1550 unique supernovae within the redshift range of $0.001 < z < 2.26$. Additionally, it incorporates 77 data points from Cepheid host galaxies at very low redshifts ($0.00122 < z < 0.01682$). 

The Pantheon+ compilation consists of 18 distinct samples, each representing data from a specific supernova survey conducted over a certain period,
	$$\chi^{2} (H_{0}\alpha )=\sum _{i=1}^{1701} \frac{[\text{$\mu_{th} $}({z_{i}}(H_{0}\alpha))-\text{$\mu_{obs}$}(z_{i})]^2}{\text{$\sigma $}(i)^2}$$
The theoretical and observed distance modulus are represented by $\mu_{th}$ and $\mu_{obs}$, respectively. $\sigma(i)$ represents the standard error of the observed values. We fitted our model's free parameters by comparing empirical and theoretical distance modulus values, $\mu_{obs}$ and $\mu_{th}$, respectively. The distance model is constructed with the formula $\mu_{th}=\mu{(D_{L})}=m_{cor}-M=5log_{10}(D_{L})+\mu_{0}$. The related term $M$ is represented by absolute magnitude, apparent magnitude, and marginalized nuisance parameters $m_{cor}$ and $\mu_{0}$.Also, $\mu_{0}$ can be obtained as $\mu_{0}=5log(H_{0}^{-1}/Mpc)+25$, and $D_{L}(z)$ is the dimensionless luminosity distance defined as $D_{L}=(1+z) \int^z_{0} \frac{1}{\zeta^{*}}d\zeta^{*}$.
Here  $\zeta^*$ denotes the change of variable defined between $0$ and $z$.

  \subsection {Baryon Acoustic Oscillation (BAO)}
We also analyze the combined baryon acoustic oscillation (BAO) dataset comprising independent measurements. This BAO dataset includes data from the SDSS Main Galaxy Sample at an effective redshift of 0.15 \cite{ref13}, the six-degree Field Galaxy Survey at an effective redshift of 0.106 \cite{ref14}, and the BOSS DR11 quasar Lyman-alpha measurement at an effective redshift of 2.4 \cite{ref15}. Additionally, we incorporate the angular diameter distances and $H(z)$ measurements from the SDSS-IV eBOSS DR14 quasar survey at effective redshifts of 0.98, 1.23, 1.52, and 1.94 \cite{ref16}, as well as the SDSS-III BOSS DR12 consensus BAO measurements of the Hubble parameter and the corresponding comoving angular diameter distances at effective redshifts of 0.38, 0.51, and 0.61 \cite{ref17}. For these two BAO datasets, we take into account the complete covariance matrix in our MCMC analyses \cite{ref18,ref19}.
 The six data points, which came from the SDSS, Wiggle Z, and 6dFGS surveys with various redshifts,

$$d_A(z_{*})=\int^{z_{*}}_{0} \frac{dz}{H(z)}$$
$$D_ v(z) =\left(\frac{d_A(z)^{2}}{H(z)}\right)^{1/3}$$ 
 The redshift at the photon decoupling epoch is shown by \textcolor{red}{$z^{*}$}, the comoving angular diameter distance is indicated by $d_A(z)$, and the dilation scale is denoted by $D_{v}$.
The $\chi^{2}$ for BAO is defined as part of the analysis,
\begin{equation}
	\label{33}
	\chi^{2}_{BAO}= X^{T}C^{-1}_{BAO} X, 
\end{equation}
where $C_{BAO}$ is the covariance matrix, $C^{-1}_{BAO}$ denotes the inverse covariance matrix, and the determination of $X$ depends on the particular survey under analysis [47]. In this case, it is 

\[
X =
\left( {\begin{array}{c c}
		\frac{d_{A}(z_{*})}{D_{v}(00.106)}& -30.95 \\
		\frac{d_{A}(z_{*})}{D_{v}(00.2)}& -17.55\\
		\frac{d_{A}(z_{*})}{D_{v}(00.35)}& -10.11\\
		\frac{d_{A}(z_{*})}{D_{v}(0.44)}& -8.44\\
		\frac{d_{A}(z_{*})}{D_{v}(00.6)}& -6.69\\
		\frac{d_{A}(z_{*})}{D_{v}(00.73)}& -5.45\\
		
\end{array} } \right)
\]

By employing the datasets discussed above, we have derived the optimal values of the parameters.The BAO data are influenced by the composition of matter in the universe.
\begin{figure}[H]
	\begin{center}
		\includegraphics[width=9cm,height=6cm, angle=0]{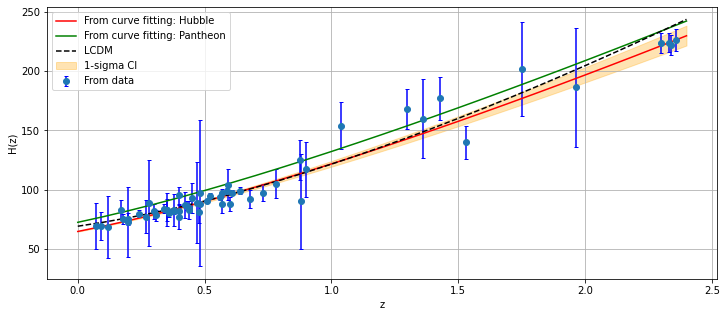}		
	\caption{The Hubble parameter error bar plot using the standard $\Lambda$CDM model and combined $CC+BAO+Pantheon+$ datasets.} 		
	\end{center}
\end{figure}
\begin{figure}[H]
	\begin{center}
		\includegraphics[width=9cm,height=6cm, angle=0]{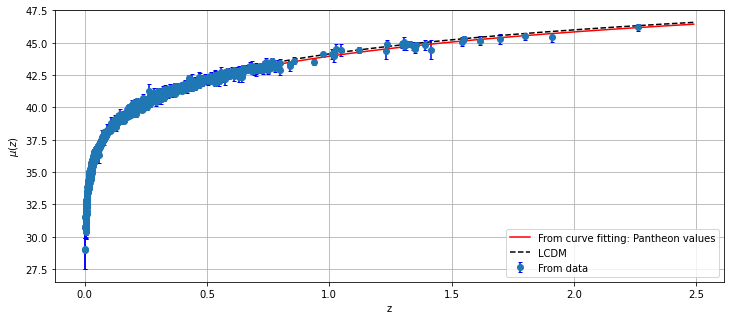}		
	\caption{ The distance modulus for the combined $CC+BAO+Pantheon+$ datasets as a function of redshift $z$.} 		
	\end{center}
\end{figure}
\begin{figure}[H]
	\begin{center}
		\includegraphics[width=9cm,height=10cm, angle=0]{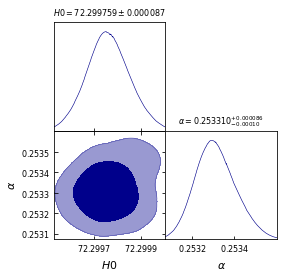}		
	\caption{The best fit plot of 1 -$\sigma$ and 2-$\sigma$ errors for the eight free parameters  for combined $CC+BAO+Pantheon+$ datasets.} 		
	\end{center}
\end{figure}
\noindent The $1-\sigma$ and 2-$\sigma$ likelihood contour for the model parameters are obtained by using the combine $CC + Pantheon+BAO$ data sets. In Fig.~1, we have presented the error bar plots using the $H(z)$ data and shown the curve pertaining our model and $\Lambda$CDM. The behavior of $\mu(z)-z$ is demonstrated in Fig.~2. Our model is fitted to 1701 points of the Pantheon+ data, along with error bars and a comparison with the $\Lambda$CDM model. Fig.~3 shows the contour graphs for the $CC+BAO+Pantheon+$ datasets at the $1-\sigma$ and $2-\sigma$ confidence levels.\\

To estimate the parameters $H_0$ and $\alpha$ along with their uncertainties, we use the Markov chain Monte Carlo (MCMC) Bayesian parameter estimation with a uniform prior. 
    After conducting various tests to ensure the independence and convergence of the MCMC samples, and appropriately thinning the samples, we derived the following estimates from the combined analysis of Pantheon data and CC.


\section{Results of observational constraints and discussion of physical behavior}

\begin{figure}[H]
 
   \includegraphics[scale=0.6]{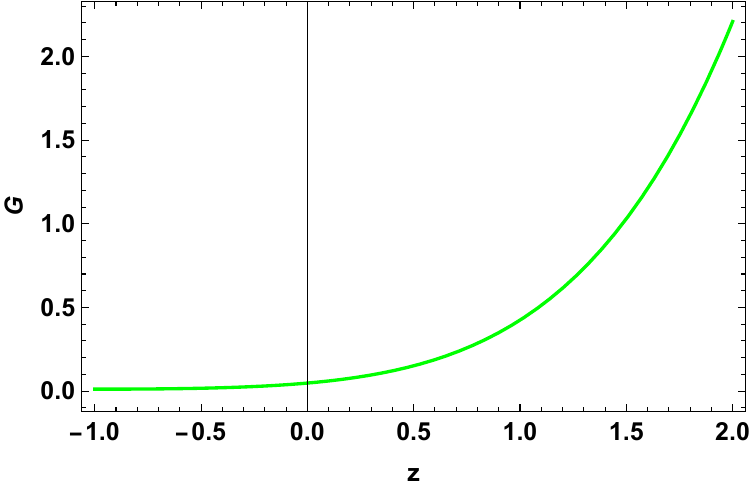}
\caption{Above figure shows the behavior of $G$ versus $z$ with the  constraint values of the cosmological free parameters limited by the combined $CC+BAO+Pantheon+$ datasets.}\label{G}
\end{figure}

The contribution of the Gauss-Bonnet term in $H$ is observed as
\begin{widetext}
\begin{equation}
G=24(2\dot{H} + H^4)=6 H_{0}^{2} \left(H_{0}^{2} \left((z+1)^{2 \alpha +2}+1\right)^2-4 (\alpha +1) (z+1)^{2 \alpha +2}\right),
\end{equation}

whereas the function with the Gauss-Bonnet invariant in $H$ is observed as
\begin{equation}
f= \alpha G^{(m + 1)} = \alpha  6^{m+1} \left(H_0^4 \left((z+1)^{2 \alpha +2}+1\right)^2-4 (\alpha +1) H_0^2 (z+1)^{2 \alpha +2}\right)^{m+1}.
\end{equation}
\end{widetext}
\noindent Fig. \ref{G} illustrates the behavior of the GB invariant $G$ as a function of redshift $z$, employing the constrained values of the model parameters. The plot indicates that the GB invariant is positive and exhibits an increasing function with redshift. Initially, at the beginning of the universe, the GB invariant starts with large positive values. However, as cosmic time progresses, it gradually approaches zero, eventually reaching this value in the far future. 

\begin{figure}[H]

   \includegraphics[scale=0.6]{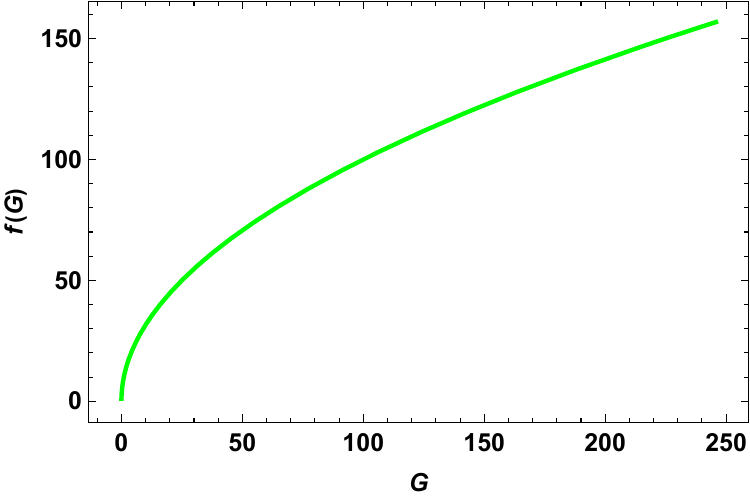}
\caption{Above figure shows the behavior of $f(G)$ versus $G$ with the  constraint values of the cosmological free parameters limited by the combined $CC+BAO+Pantheon+$ datasets.}\label{f}
\end{figure}

Similarly, in Fig. \ref{f}, we observe the behavior of the function $f(G)$ as a function of the GB invariant. The plot demonstrates a similar trend to that of the GB invariant itself, indicating a positive and crossing behavior with the GB invariant. This behavior is consistent with the dynamics expected in the model, where $f(G)$ evolves from higher values to approach zero as the universe progresses.
\begin{widetext}
The contribution of the energy density in $H$ is observed as
\[
\rho_G=\alpha  24^{m+1} \left(H^4+2 \dot{H}\right)^{m+1}-\alpha  24^{m+1} (m+1) \left(H^4+2 \dot{H}\right)^{m+1}+6 H^2+
\]
\begin{equation}
\alpha  H^4 24^{m+1} m (m+1) \left(2 \dot{H} \left(2 H^2+\dot{H}\right)+ H \ddot{H}\right) \left(H^4+2 \dot{H}\right)^{m-1},
\end{equation}
while the contribution of isotropic pressure in $H$ is observed as
\[
p_G=-6 H^2 -4 \dot{H} +\alpha  \left(-3^m\right) 8^{m+1} m \left(H^4+2 \dot{H}\right)^{m-1}\times
\]
\begin{equation}
\left(-3 H^8+8 H^6 \dot{H} (m+1)+2 H^5 \ddot{H} (m+1)+12 H^4 \dot{H} (\dot{H} m+\dot{H}-1)+2 H^3 \dot{H} \ddot{H} (m+1)+4 H^2 \dot{H}^3 (m+1)-12 \dot{H}^2\right).
\end{equation}
\end{widetext}

\begin{figure}[H]
   \includegraphics[scale=0.65]{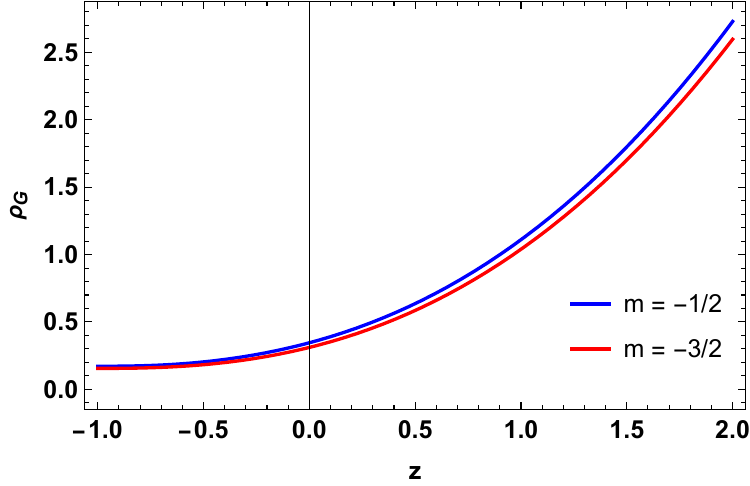}
\caption{Above figure shows the behavior of $G$ versus $z$ with the  constraint values of the cosmological free parameters limited by the combined $CC+BAO+Pantheon+$ datasets.}\label{den}
  \end{figure}

\begin{figure}[H]
    \includegraphics[scale=0.65]{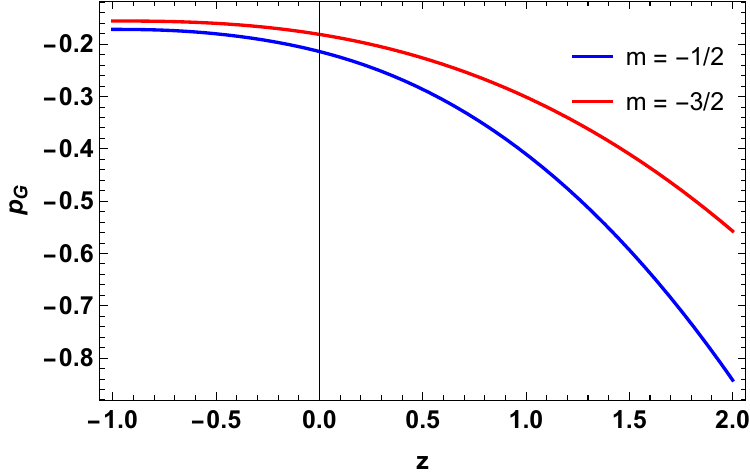}
\caption{Above figure shows the behavior of $f(G)$ versus $G$ with the  constraint values of the cosmological free parameters limited the combined $CC+BAO+Pantheon+$ datasets.}\label{p}
\end{figure}

\noindent From Fig.~\ref{den}, it is evident that the contribution of energy density $\rho_G$ exhibits the expected positive behavior, increasing with cosmic redshift, and eventually vanishing as the universe expands into the far future. This behavior holds for various values of $m$ as $m=-0.5$ and $m \ge -3/2$, except $m = -1$. On the other hand, the isotropic pressure, as presented in Fig.~\ref{p}, demonstrates a negative evolution, indicating a behavior that can drive the accelerating expansion of the universe. These trends are consistent with the expected behavior in cosmological models.

In addition, the equation of state parameter allows for the identification of other significant epochs in the universe evolution. For instance, during inflation, the equation of state parameter is approximately $\omega \approx -1$, corresponding to a period of rapid exponential expansion driven by a scalar field. In contrast, during the current epoch of cosmic acceleration, the equation of state parameter for dark energy is also close to $\omega \approx -1$, indicating a phase of accelerated expansion due to the repulsive nature of dark energy.  Also, different epochs can be characterized as follows: (a) $\omega = 1$ represents a stiff fluid; (b) $\omega = 1/3$ depicts the radiation-dominated phase; (c) $\omega = 0$ indicates the matter-dominated phase. Furthermore, the equation of state parameter can describe the cosmological constant ($\omega = -1$), quintessence ($-1 < \omega < -1/3$), or phantom era ($\omega < -1$), each representing different phases of the universe evolution characterized by the behavior of the energy density relative to the pressure. From Eqs.~(22) and (23) , the equation of state parameter is obtained as 
\begin{widetext}
\begin{multline}
	\omega_G=-\left[6 H^2+\alpha  3^m 8^{m+1} m \left(H^4+2 \dot{H}\right)^{m-1} \left(-3 H^8+8 H^6 \dot{H} (m+1)+2 H^5 \ddot{H} (m+1)+12 H^4 \dot{H} (\dot{H} m+\dot{H}-1)\right.\right.\\
	 \left.\left.+2 H^3 \dot{H} \ddot{H} (m+1)+4 H^2 \dot{H}^3 (m+1)-12 \dot{H}^2\right)+4 \dot{H}\right]/	
	\left[\alpha  24^{m+1} \left(H^4+2 \dot{H}\right)^{m+1}-\alpha  24^{m+1} (m+1) \left(H^4+2 \dot{H}\right)^{m+1}\right.\\
	\left.+6 H^2+\alpha  H^4 24^{m+1} m (m+1) \left(2 \dot{H} \left(2 H^2+\dot{H}\right)+H \ddot{H}\right) \left(H^4+2 \dot{H}\right)^{m-1}\right].
\end{multline}
\end{widetext}

\begin{figure}[H]
     \includegraphics[scale=0.65]{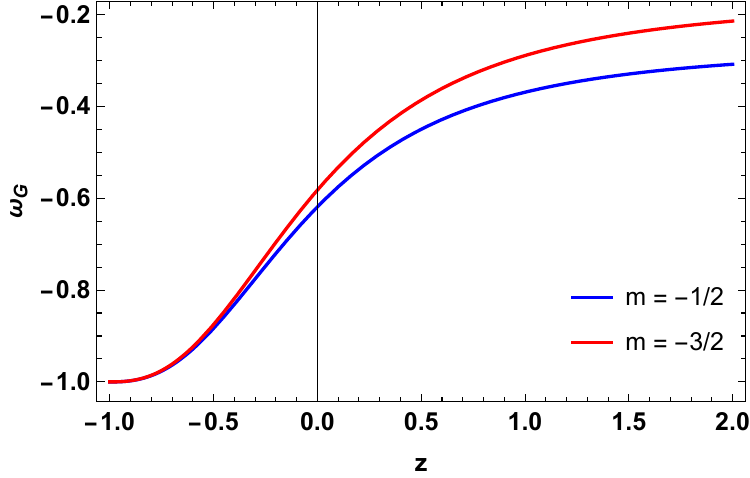}
\caption{Above figure shows the behavior of $\omega_G$ versus $z$ with the  constraint values of the cosmological free parameters limited by the combined $CC+BAO+Pantheon+$ datasets.}\label{eos}
  \end{figure}

Fig.~\ref{eos} illustrates the behavior of the parameter $\omega_G$. It is evident from the figure that $\omega_G < 0$, indicating quintessence DE and suggesting an accelerating phase. For all values of $m$, the model begins in the quintessence phase ($-1 < \omega_G < -\frac{1}{3}$) and eventually converges to a $\Lambda$CDM model ($\omega_G = -1$).

Caldwell and Linder \cite{Caldwell} have noted that the quintessence phase of dark energy can be divided into two distinct regions: thawing ($\omega_G^{\prime} > 0$, $\omega_G < 0$) and freezing ($\omega_G^{\prime} < 0$, $\omega_G < 0$) regions, delineated in the $\omega_G - \omega_G^{\prime}$ plane. Differentiating Eq.~(24) with respect to the natural logarithm of the scale factor, $a$, yields:
\begin{widetext}
\begin{multline}
	\omega_G^{\prime}=-\left[-6 H^2+\alpha  \left(-3^m\right) 8^{m+1} m \left(H^4+2 \dot{H}\right)^{m-1} \left(-3 H^8+8 H^6 \dot{H} (m+1)+2 H^5 \ddot{H} (m+1)+12 H^4 \dot{H} (\dot{H} m+\dot{H}-1)\right.\right.\\
	\left.\left.+2 H^3 \dot{H} \ddot{H} (m+1)+4 H^2 \dot{H}^3 (m+1)-12 \dot{H}^2\right)-4 \dot{H}\right]/	
	\left[(z+1) \left(\alpha  24^{m+1} \left(H^4+2 \dot{H}\right)^{m+1}-\alpha  24^{m+1} (m+1) \left(H^4+2 \dot{H}\right)^{m+1}\right.\right.\\
	\left.\left.+6 H^2+\alpha  H^4 24^{m+1} m (m+1) \left(2 \dot{H} \left(2 H^2+\dot{H}\right)+H \ddot{H}\right) \left(H^4+2 \dot{H}\right)^{m-1}\right)\right].
\end{multline}
\end{widetext}

\begin{figure}[H]
    \includegraphics[scale=0.65]{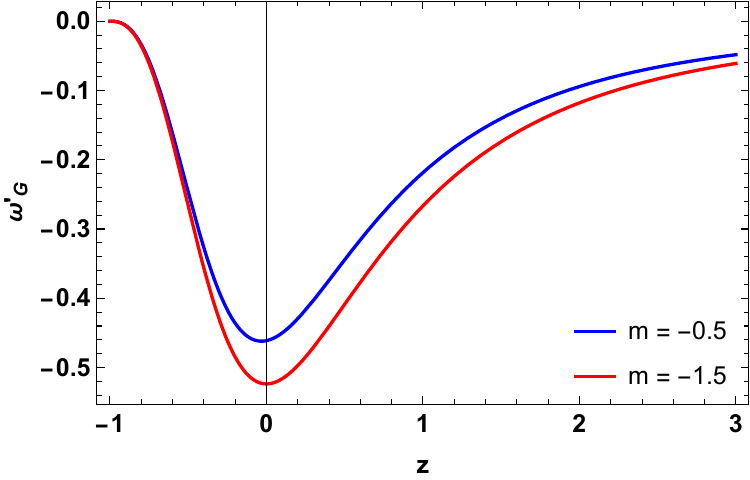}
\caption{Above figure shows the behavior of $\omega_G^{\prime}$ versus $z$ with the  constraint values of the cosmological free parameters limited by the combined $CC+BAO+Pantheon+$ datasets.}\label{w}
\end{figure}

\noindent Figs.~\ref{eos} and \ref{w} displays the $\omega_G - \omega'_G$ plane for our $f(G)$ model for the defined values of $m$. It is evident that the $\omega_G - \omega'_G$ plane corresponds to the freezing regions ($\omega_G^{\prime} < 0$, $\omega_G < 0$) for all values of $m$. This consistency indicates that the plane analysis aligns with the accelerated expansion of the Universe.

\section{Energy conditons}

Next, we will assess the viability of the obtained solution for the assumed $f(G)$ model by examining the energy conditions. The energy conditions are rules that the energy-momentum tensor must follow to guarantee the positivity of energy.  There are some well-known energy conditions named null (NEC), weak (WEC), dominant (DEC), and strong (SEC). These conditions are derived from Raychaudhuri's equation, which plays a crucial role in cosmology, and are expressed as \cite{Raychaudhuri},
\begin{itemize}
\item WEC $\Rightarrow$ $\rho_G \ge 0, \rho_G + p_G \ge 0.$
\item NEC $\Rightarrow$ $\rho_G + p_G \ge 0.$
\item DEC $\Rightarrow$ $\rho_G \ge 0, |p_G| \le \rho_G.$
\item SEC $\Rightarrow$ $\rho_G+3p_G  \ge 0.$
\end{itemize}
Thus now consuming overhead energy conditions, we can test the feasibility of our models. Also, it will be useful for us in determining our universe more realistically. The above set of energy conditions in $H$ and $\dot{H}$ are obtained as
\begin{widetext}
\begin{equation}
NEC=\alpha  H^2 3^m 8^{m+1} m (m+1) \left(H^2-2 \dot{H}\right) \left(2 \dot{H} \left(2 H^2+\dot{H}\right)+H \ddot{H}\right) \left(H^4+2 \dot{H}\right)^{m-1}-4 \dot{H}
\end{equation}

\begin{multline}
DEC=4 \left[3 H^2+\alpha  2^{3 m+1} 3^m m \left(H^4+2 \dot{H}\right)^{m-1} \left(-6 H^8+20 H^6 \dot{H} (m+1)+5 H^5 \ddot{H} (m+1)+6 H^4 \dot{H} (3 \dot{H} (m+1)-4)\right.\right. \\
 \left.\left. +2 H^3 \dot{H} \ddot{H} (m+1)+4 H^2 \dot{H}^3 (m+1)-24 \dot{H}^2\right)+\dot{H}\right]
\end{multline}

\begin{multline}
SEC=\alpha \left(-24^{m+1}\right) m \left(H^4+2 \dot{H}\right)^{m-1} \left[-2 H^8+4 H^6 \dot{H} (m+1)+H^5 \ddot{H} (m+1)+2 H^4 \dot{H} (5 \dot{H} (m+1)-4) \right.\\
\left. + 2 H^3 \dot{H} \ddot{H} (m+1)+4 H^2 \dot{H}^3 (m+1)-8 \dot{H}^2\right]-12 \left(H^2+\dot{H}\right)
\end{multline}
\end{widetext}

\begin{figure}[H]
     \centering
   \includegraphics[scale=0.65]{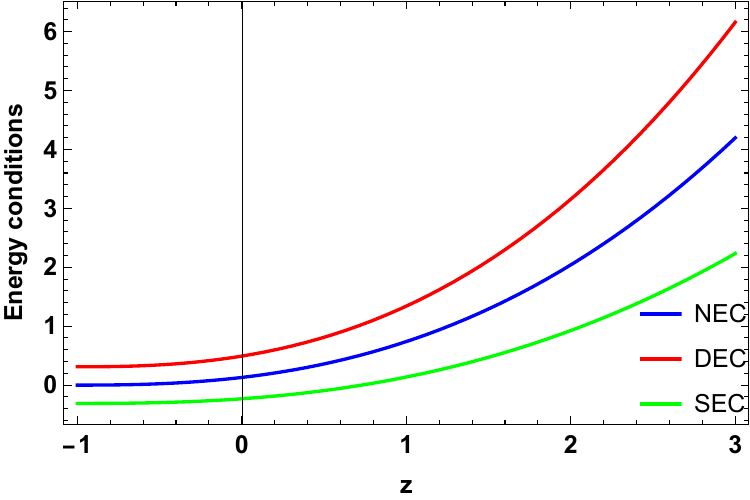}
\caption{Above figure shows the behavior of NEC ($\rho_G+p_G$), DEC ($\rho_G - p_G$) and SEC ($\rho_G + 3 p_G$) versus $z$ with the constraint values of the cosmological free parameters limited by $CC+BAO+Pantheon+$.}\label{EC}
   \end{figure}

From Fig.~\ref{EC}, it is apparent that both the NEC and the DEC are met throughout the entire range of redshifts considered, for all values of $m$. This indicates that the energy conditions required for physical viability are satisfied by the model across its entire evolutionary history. As the WEC includes both the energy density and the NEC, it is also satisfied by the model. Furthermore, Fig.~\ref{EC} illustrates that the SEC is violated in the present, for all $m$, indicating a preference for cosmic acceleration.

\section{Deceleration parameter}
Based on cosmological observations, the acceleration of the universe is a relatively recent development. To comprehensively describe the evolution of the universe, a cosmological model should account for both its decelerated and accelerated expansion phases. Therefore, it is crucial to examine the behavior of the deceleration parameter $q$ which, in terms of $H$, is written as
\begin{equation} \label{eq24}
q(z)=-1-\frac{\dot{H}}{H^2}.
\end{equation}

\begin{figure}[h]
\centerline{\includegraphics[scale=0.750]{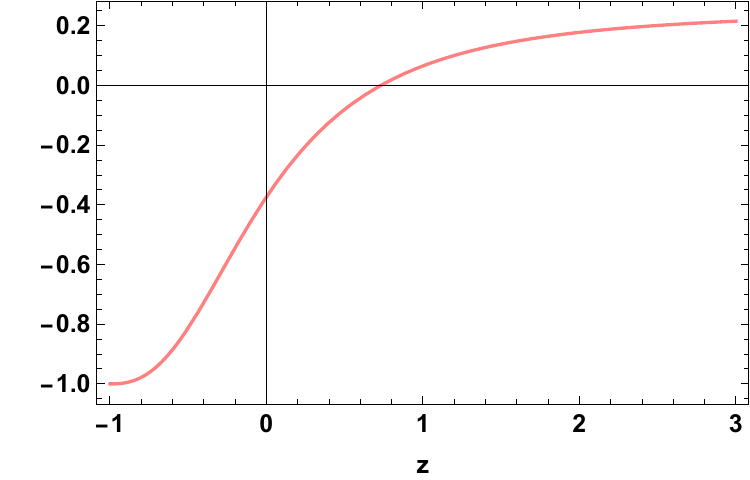}}
\caption{Above figure shows the behavior of $q$  versus $z$ with the  constraint values of the cosmological free parameters limited by$CC+BAO+Pantheon+$}\label{q}
\end{figure}

\noindent The sign of the deceleration parameter ($q$) indicates whether the model is experiencing acceleration or deceleration. A positive $q$ indicates a decelerating expansion, while $q = 0$ represents a constant rate of expansion, and $-1 < q < 0$ signifies an accelerating expansion. For $q = -1$, the universe undergoes exponential expansion, also known as de Sitter expansion, and for $q < -1$, the expansion is super-exponential. According to Fig. \ref{q}, we can observe that the evolution from a decelerating phase to an accelerating phase as redshift $z$ increases. The phase transition occurs at $z_{t} = 0.58 \pm 0.30$, a value consistent with previous literature \cite{Yadav}. Furthermore, the present value of the deceleration parameter is negative, indicating the current accelerating behavior of our universe.
\section{Statefinder analysis}

\begin{figure}[H]
\centerline{\includegraphics[scale=0.70]{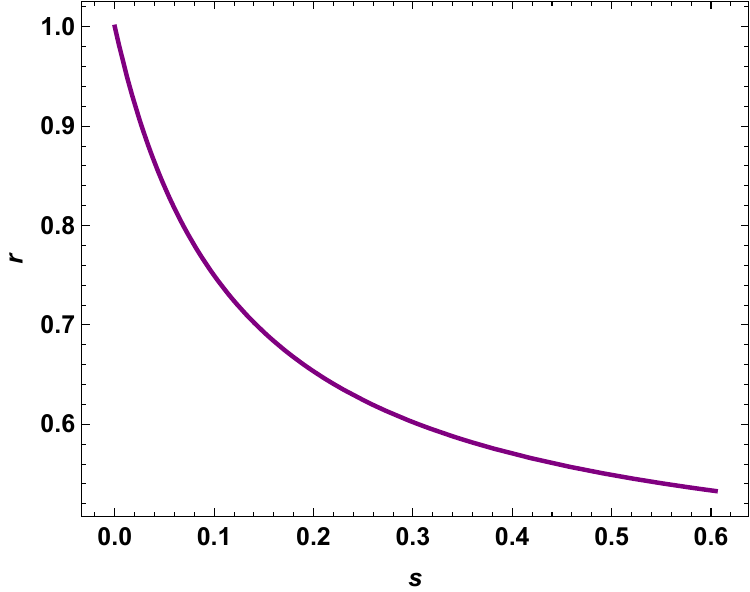}}
\caption{Figure shows the behavior of $r$-$s$ plane from the combine datasets of $CC+BAO+Pantheon+$ datasets. }\label{rs}
\end{figure}

The role of dark energy in driving the cosmic expansion is well established. Over the past few decades, there has been a growing interest in understanding the origin and fundamental properties of dark energy. As a result, numerous models of dark energy have emerged, highlighting the need to distinguish between these models, both quantitatively and qualitatively. In this regard, Sahni and et al. \cite{Sahni} introduced a statefinder diagnostic method designed to differentiate between different dark energy models. This method utilizes a pair of geometrical parameters known as statefinder parameters ($r$, $s$), defined as follows:
\begin{equation}    \label{eq26}
r= q(z)+2q^{2}(z)- H^{-1}\dot{q} ,
\end{equation} 
\begin{equation}\label{eq27}
s=\frac{(r-1)}{3(q-1/2)}  .
\end{equation} 
Different values of $(r, s)$ represent different DE models, i.e. $(r=1, s=0)$ is the $\Lambda$CDM model, $(r>1, s<0)$ is the Chaplygin Gas (CG) model, and $(r<1, s>0)$ is the Quintessence model. The graphical illustration of the statefinder diagnostics with the proper superior of constants is shown in Fig.~\ref{rs},  we see that our model's development trajectory comes to a stop at $(r,s)=(1,0)$ and $z=-1$. The behavior of $r$ and $s$ are observed at $r\rightarrow 1$ and $s \rightarrow 0$. So, our model corresponds to the $\Lambda$CDM model at late times.\\
\section{Conclusions}
In this article, we have examined the significance of the reconstruction method in driving the late-time acceleration of the universe within the framework of the $f(G)$ gravity. The present paper likely deals with the cosmological implications of modified gravity theories, specifically in the context of the $f(G)$ gravity, where $G$ represents the Gauss-Bonnet invariant. The phase-space analysis allows the identification of critical points that represent different cosmological phases (e.g. the de Sitter universe, matter-dominated era, and radiation-dominated era). Stability of these critical points is assessed to determine whether the universe would remain in those states or transition to another state. The $\Lambda$CDM model (cosmological constant $\Lambda$ + Cold Dark Matter) serves as the standard model of cosmology. In the context of the $f(G)$ gravity, the authors explore how deviations from the General Relativity (GR) can reproduce or differ from the predictions of the $\Lambda$CDM model. \\

The study is important for several reasons, particularly in advancing our understanding of the universe within the framework of modified gravity theories. The $\Lambda$CDM model, which incorporates dark energy in the form of a cosmological constant ($\Lambda$) and cold dark matter (CDM), has been highly successful in explaining large-scale structure and the evolution of the universe. However, the model faces challenges at certain scales, such as the cosmological constant problem and discrepancies in measurements of the Hubble constant. This study extends $\Lambda$CDM by considering a modified theory of gravity known as $f(G)$ gravity. Such modifications can help address issues that arise in standard $\Lambda$CDM and provide alternative explanations for cosmic acceleration. Phase-space analysis is a powerful mathematical tool that allows researchers to study the dynamical properties of cosmological models by analyzing the stability and behavior of different solutions. In this context, the study uses phase-space analysis to examine the evolution of the universe in f(G)f(G) gravity models. By investigating fixed points in phase space, researchers can identify cosmological solutions that correspond to various epochs of the universe, such as matter domination and late-time acceleration, and determine their stability. \\

The paper discusses whether $f(G)$ gravity can account for the observed accelerated expansion of the universe without invoking a cosmological constant, or how it modifies the behavior of dark energy and dark matter. The main findings can be summarized in the following points:
\begin{itemize}
	\item Section II presents our reconstruction of the field equations in the $f(G)$ gravity.
	
	\item One popular statistical method in cosmology is the Markov Chain Monte Carlo (MCMC), which is used to explore complicated model parameter spaces and provide probability distributions for cosmological parameters. In this study, the statistical analysis is done using MCMC.
	
	\item We use Supernovae observations and new data from the BAO and CC approaches to restrict our cosmological model's parameters. The data sets consist of 31 points from CC measurements, 6 points from BAO measurements, and 1701 points from the Pantheon + supernovae sample. In addition, we analyze the data using Bayesian analysis, the likelihood function from the emcee Python library, and the Markov Chain Monte Carlo (MCMC) approach.
	
	\item The Pantheon+SH0ES dataset is used, and it may be obtained from the GitHub repository. 1701 light curve observations from 1550 distinct supernovae in the redshift range of $0.001 < z < 2.26$ are included in this collection. It also includes 77 data points at very low redshifts ($0.00122 < z < 0.01682$) from Cepheid host galaxies. 
	
	\item We have also examined a pooled datasets of independent measurements for baryon acoustic oscillations (BAOs). The BOSS DR11 quasar Lyman-alpha measurement at an effective redshift of 2.4 \cite{ref15}, the six-degree Field Galaxy Survey at an effective redshift of 0.106 \cite{ref14}, and data from the SDSS Main Galaxy Sample at an effective redshift of 0.15 \cite{ref13} are all included in this BAO dataset.
	
	\item We have also combined the angular diameter distances and $H(z)$ measurements from the SDSS-IV BOSS DR14 quasar survey at effective redshifts of 0.98, 1.23, 1.52, and 1.94 \cite{ref16}, with the BAO consensus measurements of the Hubble parameter and the corresponding comoving angular diameter distances from the SDSS-III BOSS DR12 at effective redshifts of 0.38, 0.51, and 0.61 \cite{ref17}.
	
	\item
	The error bars using the $H(z)$ data and displayed the curve corresponding to our model and $\Lambda$CDM (see Fig.~1), whereas Fig.~2 illustrates how $\mu(z)-z$ behaves. Our model is fitted to 1701 points of the Pantheon+ data, along with error bars and a comparison with the $\Lambda$CDM model (Fig. 2). Fig.~3 shows the contour graphs for the $H(z) + BAO+Pantheo+$ datasets at the $1-\sigma$ and $2-\sigma$ confidence levels.

	\item  	Using the restricted values of the model parameters, Fig. \ref{G} shows how the Gause-Bonnet invariant $G$ behaves as a function of the redshift $z$. The GB invariant is shown to be positive and to have a rising function with redshift, according to the plot. The GB invariant initially has enormous positive values at the beginning of the cosmos. But cosmic time moves closer and closer to zero as it goes on, eventually arriving at this value in the remote future. Similar to this, we can see how the function $f(G)$ behaves as a function of the GB invariant in Fig.~\ref{f}. The figure exhibits a trend that is comparable to the GB invariant itself, suggesting a positive and crossing behavior. This result is in line with the dynamics predicted by the model, in which as the universe expands, $f(G)$ evolves from higher values to become closer to zero. 
	
	\item From Fig.~\ref{den}, it is evident that the contribution of energy density $\rho_G$ exhibits the expected positive behavior, increasing with cosmic redshift, and eventually vanishing as the universe expands into the far future. This behavior holds for various values of $m$ as $m=-0.5$ and $m \ge -3/2$, except $m = -1$. On the other hand, the isotropic pressure, as presented in Fig.~\ref{p}, demonstrates a negative evolution, indicating a behavior that can drive the accelerating expansion of the universe. These trends are consistent with the expected behavior in cosmological models.
	
	\item In conclusion, our analysis of the equation of state parameter has revealed significant insights into the evolution of the universe, identifying distinct epochs characterized by various values of $m$. We have shown that our model begins in the quintessence phase and converges to the $\Lambda$CDM model, aligning with the accelerated expansion of the universe. The $\omega_G - \omega'_G$ plane analysis confirms the freezing regions, indicating consistency with the accelerating expansion. We have also observed a phase transition from deceleration to acceleration at $z{t} = 0.58 \pm 0.30$, consistent with previous literature. The model satisfies the necessary energy conditions (NEC, DEC, and WEC) across its entire evolutionary history, while violating the SEC, indicating a preference for cosmic acceleration. Ultimately, our model's trajectory converges to the $\Lambda$CDM model at late times, providing a comprehensive understanding of the universe's evolution.
		
	\item The paper is concluded by summarizing the key findings of the phase-space analysis, including which models within the $f(G)$ framework are consistent with the known cosmological observations.
	
	\item It might also discuss the conditions under which the $f(G)$ gravity can recover the $\Lambda$CDM model or provide alternative explanations for the accelerated expansion of the universe.
	
	\item Potential implications for future research or observational tests that could distinguish the $f(G)$ gravity from the standard $\Lambda$CDM model might be suggested.
	
	The study of phase-space analysis in the $\Lambda$CDM $f(G)$ cosmology is crucial for advancing our understanding of the universe dynamics, exploring alternatives to the standard cosmological model, and potentially uncovering new physics beyond General Relativity. Phase-space analysis provides a powerful mathematical framework to study the evolution of the universe in different cosmological models. In future, by identifying critical points in the phase space, we can determine the possible past, present, and future states of the universe under the $f(G)$ gravity. This study can guide future theoretical work in cosmology, including the development of new models or the refinement of existing ones, as well as suggest new observational tests that could distinguish between different models.

\end{itemize}

\section*{Declaration of competing interest}
The authors declare that they have no known competing financial interests or personal relationships that could have appeared to influence the work reported in this paper.

\section*{Data availability}
No data was used for the research described in the article.

\section*{Acknowledgments}
This research is funded by the Science Committee of the Ministry of Science and Higher Education of the Republic of Kazakhstan (Grant No. AP23483654). The authors, A. Pradhan \& S. H. Shekh, also thank the IUCAA, Pune, India, for providing the facility through the Visiting Associateship programs.

\end{document}